\documentclass[aps,prd,preprintnumbers,showpacs,nofootinbib,twocolumn]{revtex4}
\usepackage{graphicx}
\usepackage{epsfig}
\usepackage{amsmath, amssymb}
\usepackage{dcolumn}
\usepackage{bm}
\usepackage{amsfonts}

\usepackage{subfigure}
\usepackage{amssymb}
\usepackage{afterpage}

\def\dj{\hbox{d\kern-0.347em \vrule width 0.3em height 1.252ex depth
-1.21ex \kern 0.051em}}

\newcommand{\be}{\begin{equation}}
\newcommand{\ee}{\end{equation}}
\newcommand{\ben}{\begin{equation*}}
\newcommand{\een}{\end{equation*}}
\newcommand{\bea}{\begin{eqnarray}}
\newcommand{\eea}{\end{eqnarray}}
\newcommand{\bean}{\begin{eqnarray*}}
\newcommand{\eean}{\end{eqnarray*}}
\newcommand{\brr}{\begin{array}}
\newcommand{\err}{\end{array}}
\newcommand{\bc}{\begin{center}}
\newcommand{\ec}{\end{center}}

\newcommand{\lsim}{\,\raisebox{-0.6ex}{$\buildrel < \over \sim$}\,}
\newcommand{\gsim}{\,\raisebox{-0.6ex}{$\buildrel > \over \sim$}\,}

\newcommand{\bk}{{\mathbf k}}

\newcommand{\by}{{\mathbf y}}
\newcommand{\bx}{{\mathbf x}}

\newcommand{\de}{\delta}

\newcommand{\La}{\Lambda}

\newcommand{\Om}{\Omega}

\newcommand{\si}{\sigma}

\begin{document}

\title{Effects of biasing on the galaxy power spectrum at large scales}
\author{Jos\'e Beltr\'an Jim\'enez$^{1,2}$ and Ruth Durrer$^{1}$}
\affiliation{$^1$Institute de Physique Th\'eorique, Universit\'e de Gen\`eve, 24 quai 
E. Ansermet, 1211 Gen\`eve 4, Switzerland\\
$^2$ Departamento de F\'isica Te\'orica, Universidad Complutense 
de Madrid, 28040, Madrid, Spain. }


\date{\today}

\begin{abstract}In this paper we study the effect of biasing on the power
spectrum at large scales. We show that even though non-linear biasing does 
introduce a white noise contribution on large scales, the $P(k)\propto k^n$
behavior of the matter power spectrum on large scales may
still be visible and above the white noise for about one decade.
We show, that the Kaiser biasing scheme which leads to linear bias of the 
correlation function on {\em large } scales, also generates a linear bias of the
{\rm power spectrum} on rather small scales. This is a consequence of the divergence 
on small scales of the pure Harrison-Zeldovich spectrum. However, 
biasing becomes k-dependent when we damp the underlying power 
spectrum on small scales.
We also discuss the effect of biasing on the baryon 
acoustic oscillations.
\end{abstract}

\maketitle

\section{Introduction}\label{introduction}

One of most promising future observations for cosmology is the precise 
determination of the galaxy power spectrum. So far, the emphasis has been on 
the anisotropies and polarization of the cosmic microwave background (CMB)
but in the future we also want to determine the matter distribution of the 
Universe with much better precision. The advantage of the matter distribution
if compared to the CMB is that while the latter represents only a 
two-dimensional data set, it has been emitted from the surface of last 
scattering, the former is three-dimensional and therefore contains, in 
principle much more information. The disadvantage is that we can only observe 
galaxies and it is not clear how the observed galaxy distribution is 
related to the underlying matter distribution which we calculate using 
cosmological perturbation theory~(see e.g. \cite{mybook}) and numerical N-body 
simulations~(see e.g. \cite{springel}). 

This problem goes under the name of 'biasing' and has been studied in many
works. Among the first influential papers on the topic 
are~\cite{kaiser,poli,bbks}. They are based on the idea that galaxies form
at the peaks of the matter distribution or above a certain threshold. A
first interpretation of the 'Kaiser biasing model' was that bias is linear
at least on large scales~\cite{kaiser}. But data have shown that bias is not 
linear, see e.g.~\cite{24,25}, and this is actually also the case in the full 
'Kaiser biasing model'~\cite{22,sylos}. Subsequently, a fitting formula for 
bias has been proposed~\cite{26}, which is often used, but is not derived from
physical considerations and leads to cosmological parameters which are not 
in good  agreement with other observations~\cite{27}. Most modern versions of 
biasing have replaced the 'peaks' of the density distribution by so called 
'dark matter halos' inside which galaxies are supposed to 
form~\cite{73,71,67,68,65,61}. Even though these models do qualitatively
agree with observations~\cite{79}, they are very model dependent. One has
to specify a so called halo occupation distribution (HOD). Even simple 
models of this have of the order 10 free parameters~\cite{63}. Furthermore, 
simulations suggest that the HOD probably depends on time~\cite{89,90} and on 
the details of the formation process (e.g. whether the halo under consideration 
has been ejected from another halo~\cite{93} or undergone mergers~\cite{91}).
An alternative idea, where  galaxies are considered as the fundamental objects and 
halos are attached to them is put forward in Ref.~\cite{66}. 

Another modern approach to biasing is perturbation theory. It is based on 
the assumption that the galaxy distribution can be expanded in a Taylor 
series of the density fluctuation at the same point (local 
bias)~\cite{99,vinc}. This can be generalized by allowing also a dependence 
on the divergence of the velocity field and on shear~\cite{McDRoy}. The 
advantage of this approach is that on
large scales, where fluctuations are small, one can reduce it to only a
few free parameters. The disadvantage is that one has no a priori information
on the amplitude of these unknown parameters and one has to fit them together 
with other cosmological parameters. Furthermore, the white noise contribution
in the power spectrum on large scales is also present in this biasing scheme
as soon as one allows for non-linear contributions. This has
already been noticed in~\cite{100,sylos}. In addition, the results depend
on whether one uses Lagrangian or Euclidean perturbation theory~\cite{117,116}.
A systematic comparision between higher order Eulerian and Langrangian 
perturbation theory with numerical simulations can be found in~\cite{EvsL}.

In this paper we do not develop any detailed biasing model, but we want to 
address two main issues which we believe are quite model independent. First, 
we want to illustrate the fact that even though the galaxy 
correlation function may differ from the matter correlation function by more 
than a simple multiplicative constant (non-linear biasing) only on small 
scales, this can significantly modify the galaxy power spectrum also on 
very large scales.

Secondly we want to estimate the effects of biasing on intermediate
scales which are relevant for
baryon acoustic oscillations (BAO's). These are the remains of the oscillations 
in the baryon-photon plasma generated prior to recombination which then 
contribute to the matter power spectrum. These acoustic peaks have been 
measured in detail in the CMB anisotropy spectrum~\cite{WMAP7}, and there is
 evidence of their presence also in the galaxy power 
spectrum~\cite{BAOs}.  Even though there are some doubts about the reality of these
features~ \cite{NOBAO}, there are now new claims of the detection of BAO's even 
above 4-sigma~\cite{4sigma}. In any case, substantial progress is expected with future surveys like 
DES~\cite{DESS}, BOSS~\cite{BOSS} or the satellite project 
Euclid~\cite{EUCLID}.

In this brief paper, we are not concerned to use the most sophisticated or 
realistic model of biasing, since we mainly want to make a point of principle. 
In subsequent work we shall investigate the ideas presented here in detail, 
using a more realistic 
biasing scheme e.g. the one studied in Ref.~\cite{vinc}. 

In the next section we determine the biased correlation function and power 
spectrum for a simple non-linear biasing model and discuss the main features.
In section~\ref{s:con} we conclude.

\section{The biased power spectrum}
Non-linear Newtonian clustering does not induce an inverse cascade, but rather 
moves power from large to smaller scales (direct cascade). 
This effect flattens the density power spectrum on small scales which
according to linear perturbation theory should behave like 
$P(k)\propto k^{-3}\ln^2(k)$ to a behavior roughly like $P(k)\approx k^{-2}$ 
found in numerical simulations, see e.g.~\cite{springel}.
 On large scale, inflation predicts that $P(k)\propto k^{n}$,
and the spectral index $n= 0.963\pm0.014$ has been measured with the WMAP
satellite~\cite{WMAP7}. This is a very special power spectrum with $P(0)=0$.
Since the power spectrum is the Fourier transform of the correlation function,
\be\label{e:Pxi}
P(k) = \sqrt{\frac{2}{\pi}}\int_0^\infty \xi(r)j_0(kr)r^2dr\, , \quad 
 j_0(x) =\frac{\sin x}{x} \, ,
\ee
\be\label{e:intc}
P(0) =0~\quad \mbox{ implies }\quad \int_0^\infty \xi(r)r^2dr =0 \, .
\ee
This integral constraint is very unusual and implies that the matter 
distribution represents a 'super-homogeneous' system in a precise statistical 
sense~\cite{sylos2,sylosBook}.

For example, this requires that there is absolutely no white noise in the 
system because this would add a constant to the power spectrum. From 
Eq.~(\ref{e:intc}) it is clear that if biasing is not everywhere
linear, leading to a galaxy correlation function which is not simply 
$\xi_g(r) = b\xi(r)$, there is a high chance that the very subtle and non 
local integral constraint~(\ref{e:intc}) will be violated and
$P_g(0)\neq 0$. This actually happens in {\em all} the non-linear biasing 
models mentioned in the introduction. Of course, if one 
subtracts the mean galaxy density in an observation to obtain $\de(\bx)$ one has, by
construction $\de(\bk=0)=0$ and hence also $P(k=0)=\langle|\de(0)|^2\rangle =0$. 
However, one, in
principle has to test whether this mean density is well defined in the sense
that it is the same in half the observation volume as in the entire volume.
If this is not the case, one has to be cautious of the fact that one is subtracting 
a fictitious mean density which is not truly well defined. In this sense, here by
$P(k=0)$ we actually mean the theoretical infinite volume limit  $\lim_{k\rightarrow 0} P(k)$
in which the power spectrum of a sample with no correlations (Poissonian) tends to
a non-zero constant (white noise).

We illustrate the point that the biased power spectrum acquires such a 'white noise'
contribution by using the simple biasing scheme which has been 
proposed by Kaiser~\cite{kaiser}: be $\de(\bx)$ the density fluctuations 
with variance $\langle \de(\bx)^2\rangle=\sigma^2 =\xi(0)$. We assume that 
galaxies form when the density fluctuation is larger than a threshold $\nu$, 
$\de(\bx)>\nu\si$. The correlation function of galaxies forming according to 
this prescription is given by the correlation function of the threshold sets
$\theta_\nu$ defined by~\cite{kaiser,poli}
\be\label{3}
 \theta_\nu(\bx) \equiv \theta(\de(\bx)-\nu\si)= \left\{ \begin{array}{ll}
	1 & \mbox{if }~~~ \de(\bx) \ge \nu\si \\
	0 & \mbox{else.}
\end{array} \right.
\ee
$\langle \theta(\bx)\rangle=\langle \theta(\bx)^2\rangle=Q(\nu)$ gives the 
fraction of the volume in which $\de(\bx)>\nu\si$.  Note that this biasing 
scheme is local, $\theta_\nu(\bx)$ depends only on $\de(\bx)$, but, as we
shall see, it is non-linear. By construction, the
amplitude of $\theta_\nu$ never exceeds $1$ and therefore also its correlation 
function,
$$ \xi_\nu(r) = \langle \theta_\nu(\bx) \theta_\nu(\by)\rangle \le 1  \qquad
 \forall \quad r=|\bx-\by| \,.$$
This overall normalization is arbitrary and so is the overall normalization
of the corresponding power spectrum $P_\nu$. We shall therefore not comment 
on the overall amplitude of the power spectrum in the following. It has no physical 
significance and could e.g. be multiplied with $\si^2$ or $(\nu\si)^2$.

If we assume that $\de(\bx)$ is a 
Gaussian field with vanishing mean, the one point distribution is
\[ P(\de) = \frac{1}{\sqrt{2\pi}\si} e^{-{\de^2\over 2\si^2}} ~,\] 
and the 2-point function is given by
\be P(\de_1,\de_2, r) = \frac{ \exp\left( -\frac{\si^2(\de_1^2+\de_2^2) -2\xi(r)\de_1\de_2}{
2(\si^4-\xi^2(r))}\right) \,}{2\pi\sqrt{\si^4-\xi(r)^2}}
 ,
\ee 
where $\de_1=\de(\bx_1)$, $\de_2=\de(\bx_2)$ and $r =|\bx_1-\bx_2|$.
From this it is easy to derive~\cite{kaiser,sylos} that the correlation function 
for the biased field $\theta_\nu$ is given by
 \be
\xi_\nu(r) =\frac{\int_\nu^\infty dx e^{-x^2/2}\int_{\mu(r)}^\nu dye^{-y^2/2}}{
              [\int_\nu^\infty dx e^{-x^2/2}]^2} \,,
\ee
where $\mu(r)=(\nu\si^2-\xi(r)x)/\sqrt{\si^4-\xi^2(r)}$.
Clearly, if $\xi(r)=0$, $\mu(r)=\nu$ and hence also $\xi_\nu(r)=0$, both
$\xi$ and $\xi_\nu$ pass through zero at the same scale $r_0$. Furthermore,
it can be shown that on large scales, where $\xi(r)\ll \si^2$ and 
$\xi_\nu(r) \ll 1$ the correlation functions satisfy the linear relation 
$\xi_\nu(r) \simeq \nu^2\xi(r)$. For this reason, this biasing scheme is often 
called 'linear bias'. However, as has been pointed out in 
Ref.~\cite{sylos}, the biased power spectrum,
\be
P_\nu(k) =\sqrt{\frac{2}{\pi}}\int_0^\infty \xi_\nu(r)j_0(kr)r^2dr
\ee
does no longer vanish at the origin,
\be
P_\nu(0) =\sqrt{\frac{2}{\pi}}\int_0^\infty \xi_\nu(r)r^2dr \neq 0 \,.
\ee
Even though $\xi_\nu(r) = \nu^2\xi(r)$ on large scales, we do not have
$P_\nu(k)=\nu^2P(k)$ for small $k$.
We now want to estimate the amplitude $P_\nu(0)$ in order to decide whether 
the turn over to the $P(k)\propto k^n$ behavior can be seen in the galaxy 
power spectrum or it is completely masked by this biasing effect.
Note also, that we have not introduced any ad hoc white noise component,
 but just violated the super-homogeneity condition (\ref{e:intc}). As 
mentioned above, this happens for all generic non-linear biasing schemes.

\subsection{Very large scales}

We have calculated the biased power spectrum $P_\nu(k)$ from an underlying 
$\La$CDM matter
power spectrum as approximated in~\cite{dod} for different values of the
biasing parameter $\nu$. This calculation has already 
been performed in Ref.~\cite{sylos}, but there an exponential cutoff
on small scales has been introduced  for convenience. The reason for using such a 
cutoff was  to improve the convergence properties of the integrals involved in the 
computation of the biased power spectrum given the underlying one. However, as we 
shall see, such a cutoff can hide some effects coming from the small scales that do 
impact on the results obtained in Ref.~\cite{sylos}. Here we use the full spectrum given 
in Ref.~\cite{dod} which does, however neglect baryons and makes use of the fitting 
formula for the transfer function obtained in \cite{bbks}. Also, this power spectrum neglects
non-linearities which are in principle relevant on small scales. The linear
CDM transfer function is given by~\cite{bbks}:
\bea
T(x)=\frac{\ln(1+0.171x)}{0.171x}\left[1+0.284x+(1.18x)^2\right.\nonumber\\
\left.+(0.399x)^3+(0.490x)^4\right]^{-1/4}
\eea
with $x\equiv k/k_{eq}$ and $k_{eq}=\sqrt{2(1+z_{eq})\Omega_M}H_0$.
The matter power spectrum is given in terms of the transfer function by 
$P(k)=T^2(k)P_0(k)$, with $P_0(k)\propto k^n$ the primordial power spectrum. This fitting 
formula does not contain non-linear correction to the matter power spectrum that would enhance the small scale power and, with it, $\si^2$.
This would  therefore even enhance the difference between the result presented here 
and in Ref.~\cite{sylos}.  

\begin{figure*}
{ \epsfxsize=17cm \epsfbox{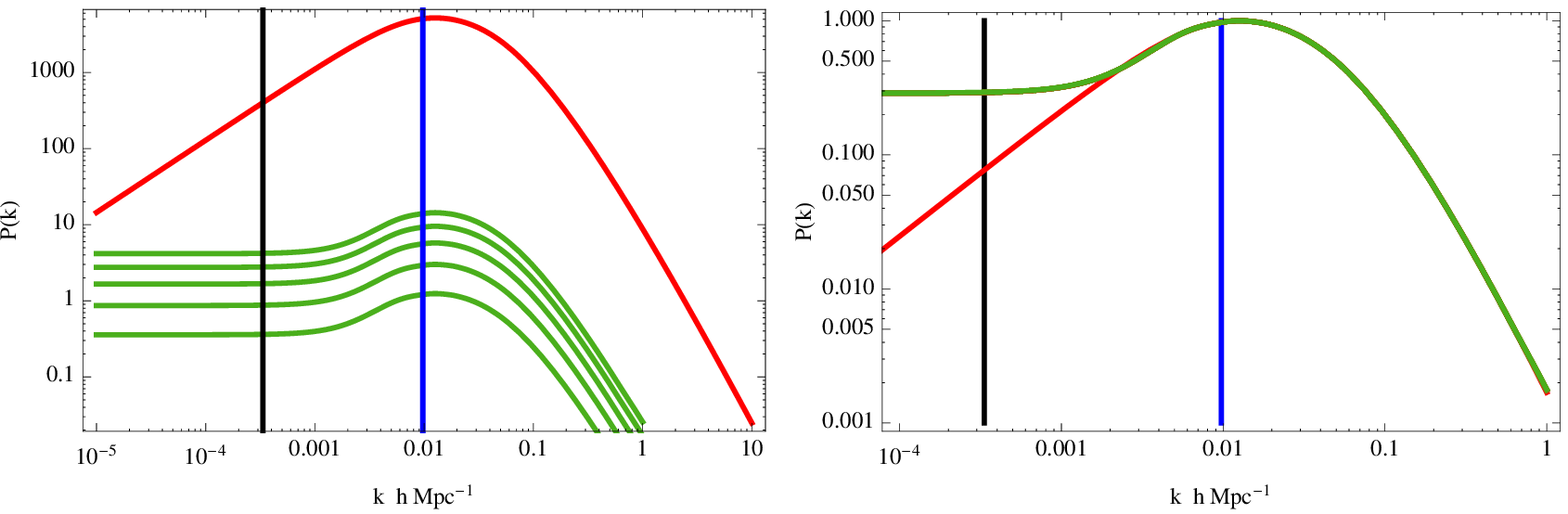}}
\caption{In the left panel we show the underlying $\La$CDM power spectrum 
(red line) and the corresponding biased power spectra (green lines) for 
$\nu=5$, $4$, $3$, $2$ and $1$ from top to bottom. As mentioned in the text, the amplitude of
the biased power spectra is unphysical. In the right panel we 
have normalized the power spectra to their maximum value after which they 
all collapse to $P_1(k)$ as discussed in the main text.  
We have also indicated the scales corresponding to the horizon scale at 
equality, $k_{eq}$ (blue vertical line), and the present horizon scale, 
$k_0$ (black vertical line).}
\label{PBBKS}

\epsfig{width=17cm, file=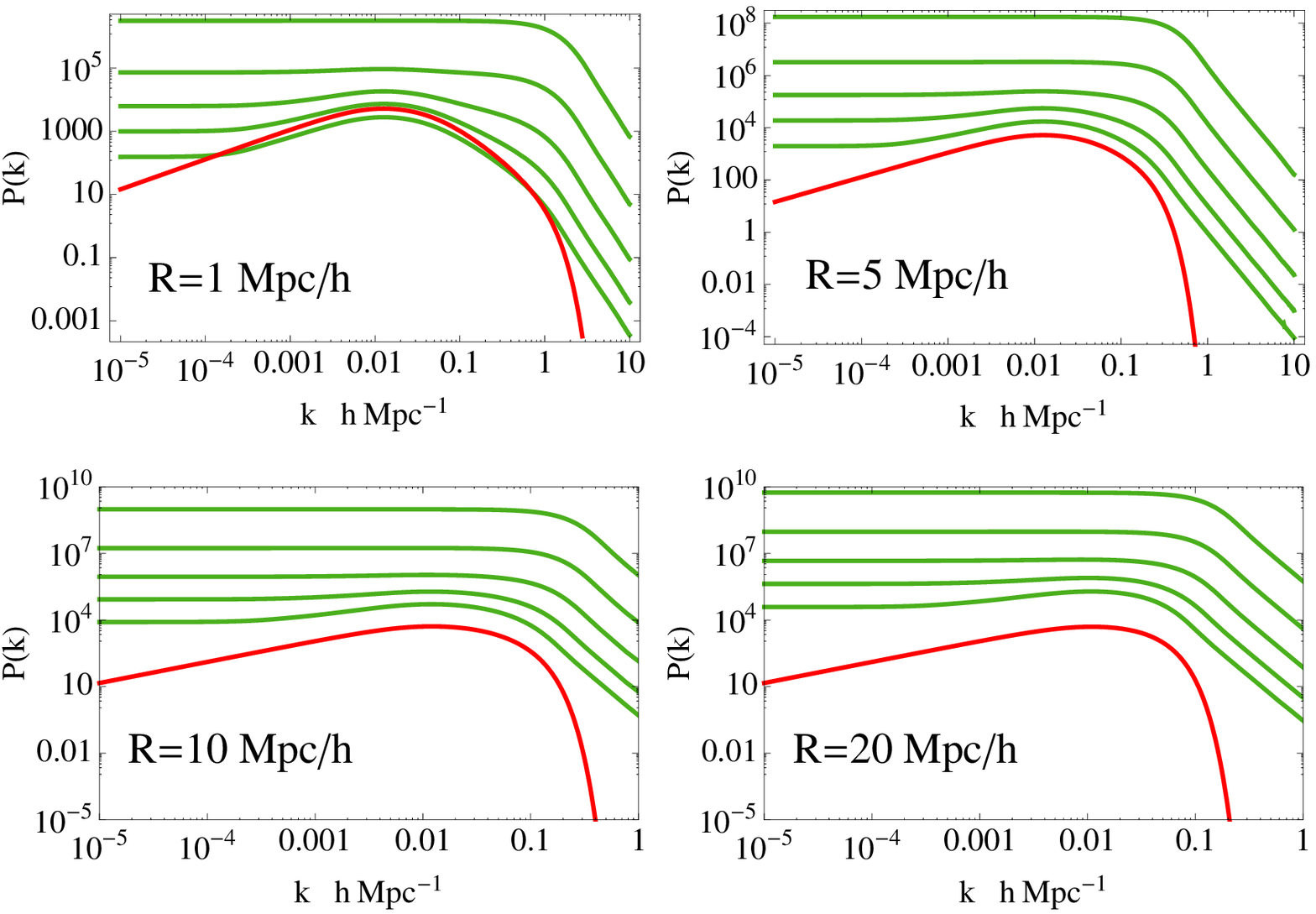}
\caption{We show the biased power spectrum for a smoothed underlying power 
spectrum as explained in the text for several smoothing scales (indicated in 
the corresponding panels) and for $\nu=$5, 4, 3, 2 and 1 from top to bottom. Note also here that
the amplitude of the biased power spectra is unphysical. Actually, the value $P(0)$ 
is (up to a factor $\nu^2$) the fraction of the volume for which $\xi_c(r)=\xi(r)/\sigma^2$ 
is larger than $\nu$ which becomes large for larger damping scale $R$ since damping reduces 
the variance $\si^2$.}
\label{PSR}
\end{figure*}

In Fig.~\ref{PBBKS} we show the biased power spectrum obtained from the
underlying  $\La$CDM matter power spectrum. One clearly sees that, even though
the underlying spectrum behaves like $k^n$ at large scales (small $k$), the biased spectra tend to a constant since the 
integral constraint (\ref{e:intc}) is violated, i.e., there is no longer an exact 
cancellation between correlations at small scales and anti-correlations at large scales. It is 
interesting to note that the modification of the power spectrum can be split 
into two effects: a universal modification in the shape and a vertical 
shift (a reduction of power in the present case) that depends on the 
biasing parameter $\nu$. This becomes apparent when we normalize the biased 
spectra to their maximum value. As shown in Fig \ref{PBBKS}, the normalized 
biased power spectra share the same shape irrespectively of the particular 
value of the parameter $\nu$. Moreover, we see that the only modification 
when compared to the underlying $\La$CDM spectrum is the appearance of the 
plateau for very large scales, spoiling the $P(k)\propto k^n$ behaviour. 
Thus, we can conclude that the biased power spectrum can be factorized 
as $P_\nu(k)=b_1(\nu)P_1(k)$, with $b_1(\nu)$ accounting for the 
$\nu$-dependent shift  and $P_1(k)$ the universal biased power spectrum 
shown in Fig.~\ref{PBBKS} that becomes a constant for large scales and
tends to the underlying power spectrum $P(k)$ on small scales.
We obtain the following 
analytical expression for the shift function:
\be
b_1(\nu)=\frac{\pi}{2\sigma^2}\frac{e^{-\nu^2}}{\left(1-{\rm erf}(\nu/
\sqrt{2})\right)^2}
\label{b1}
\ee
where ${\rm erf}(x)$ denotes the error function.
In Ref.~\cite{sylos} it is derived that $b_1(\nu)\xi(r) \simeq \xi_{\nu}(r)$ in
the regime where $\nu\xi(r)<1$ and $\xi(r)/\si^2=\xi_c(r)\ll 1$, hence on sufficiently 
large scales. Here we see that $b_1(\nu)P(k)$ is a good approximation to 
the biased power spectrum, $P_\nu(k)$ on small scales as well. This 
approximation is good for $k>k_{\rm eq}$ up to $\xi_c(1/k)\ll 1$, hence up to 
about $k\lsim 10^5$Mpc$^{-1}$.

\begin{figure}[hb!]
\begin{center}
 \epsfig{width=8cm, file=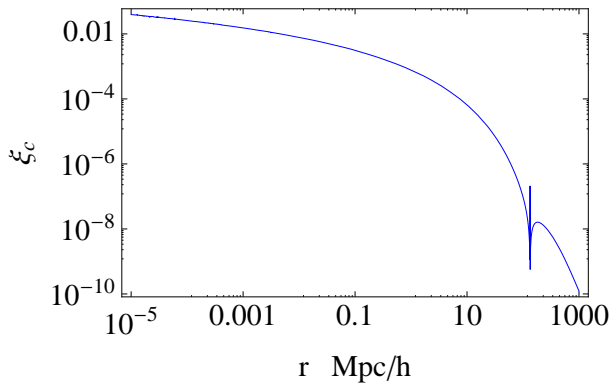}
\caption{The normalized correlation function $\xi_c=\xi/\sigma^2$ is shown. Since
the spectral index is close to one, $\sigma^2$ is very large and $\xi_c(r)$ is small
already on small scales.}
\label{xic}
\end{center}
\end{figure}

In the right panel of Fig.~\ref{PBBKS}, we compare the normalized biased power 
spectra which all collapse on one line with the underlying $\La$CDM power
spectrum. Contrary to the standard belief, the Kaiser biasing scheme leads 
to linear bias of the power spectrum on small scales and non-linear bias 
on large scales. This might seem to contradict the well-known fact that 
biasing is non-linear on small scales, as commented in the introduction. 
However, this is not the case. The reason why we obtain the linear biasing on 
small scales here is the log-divergence of the pure Harrison-Zel'dovich spectrum 
on small scales which leads to an infinite value of the correlation function 
at the origin, i.e., $\sigma^2=\xi(0)=\infty$. Notice that
\begin{equation}
\sigma^2=\xi(0)=\int_0^\infty k^2 P(k) dk.
\end{equation}
For large scales (small k) we have that $P(k)\propto k^n$ so we have no 
divergences in the lower limit, but for small scales (large k) we have 
$P(k)\propto (\log k)^2 k^{n-4}$ so the integral has a logarithmic divergence 
in the upper limit for $n=1$. We are considering a spectral index slightly 
smaller than 1, hence $\sigma^2$ is not infinity, but very large so that 
$\xi_c(r)=\xi(r)/\sigma^2$ is small already for relatively small scales,
see Fig.~\ref{xic}, and the linear approximation $\xi_\nu\simeq b_1(\nu)\xi(r)$
is valid.

Of course, on very small scales even the underlying CDM power spectrum is damped.
Typically, the CDM damping scale is expected to be very small, below 1pc. However, as 
we shall show below, once we introduce a somewhat larger damping scale which might 
relevant e.g. if a warm dark matter component is present or might come from Silk 
damping of baryons, the 
divergence of the HZ spectrum is regularized, $\sigma$ is substantially reduced, and 
the amplification of the spectrum on small scales is no longer linear.
  
\begin{figure*}
{ \epsfxsize=17cm \epsfbox{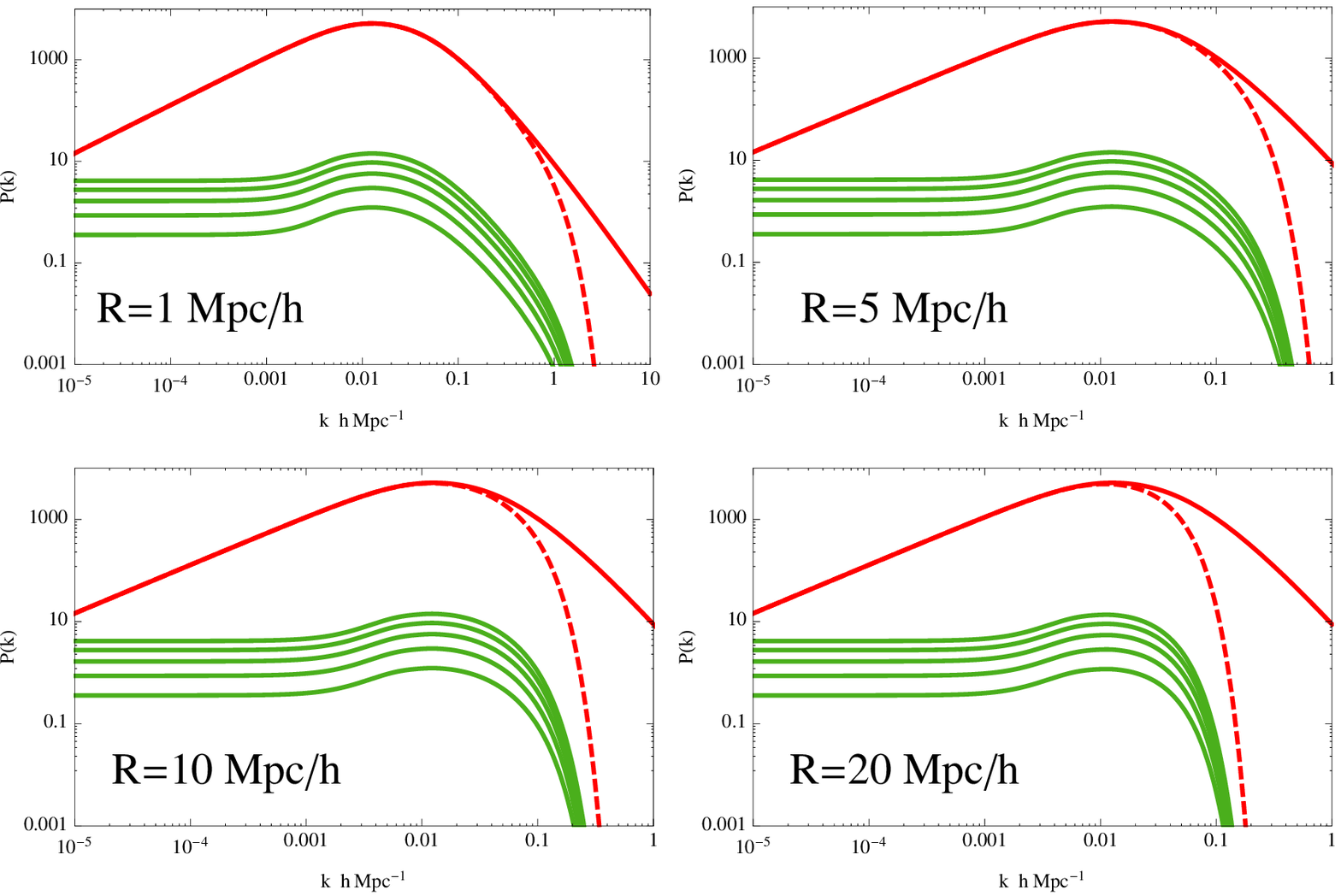}}
\caption{ In this plot we show the biased power spectra after small scale damping. The red line corresponds to the underlying power spectrum, the red-dashed line is the smoothed underlying power spectrum and the green lines show the biased power spectra after applying smoothing for $\nu=5,4,3,2,1$ from top to bottom. On large scales, $k<1/R$, the biased spectra are
not affected and correspond to the ones shown in Fig.~\protect\ref{PBBKS}}
\label{Pnusmoothed}
\end{figure*}

We now shall consider the case where small scale power has been removed from the 
underlying power spectrum. To that end, we simply smooth the underlying power spectrum
with a Gaussian window function, 
\be
P=P(k,R)=|\widehat W(kR)|^2P(k)
\ee
where $\widehat W(kR)=\exp(-R^2k^2/2)$ is the  Fourier transform of the 
normalized window function  $W(\bx,R)=\frac{1}{V_R}\exp[-\bx^2/(2R^2)]$. 

This allows us to compare with the previous results obtained in Ref~\cite{sylos},  
since the cutoff introduced
there is equivalent to the damping introduced here. Also, this might be relevant for 
models with warm  dark matter where the primordial power spectrum is damped 
on small scales. (It is not clear whether this small scale damping survives nonlinear 
effects which are also relevant on small scales and which are neglected in our treatment.)  Otherwise, this may simple be
interpreted as an alternative biasing scheme: apply Kaiser biasing not to the true 
power spectrum but to the one from which the smallest scale structure which is 
responsible for the very high value of $\si$ is removed. In 
this case, the shape of the biased power spectrum is no longer universal.
It is amplified non-linearly both, on scales smaller than the smoothing scale and
on large scales (see Fig. \ref{PSR}). Indeed, for high values of $\nu$ or large
smoothing scale, the turnover that is present in the underlying power spectrum 
disappears completely. This disappearance happens for smaller values of $\nu$ 
as the  smoothing scale increases. These results are in line with those 
obtained in Ref.~\cite{sylos} since, as aforementioned,  the smoothing applied here is 
equivalent to the exponential cutoff for small scales introduced in that work. As 
explained above, once we introduce smoothing over a scale $R$, power on scales 
smaller than $R$ is damped and the UV divergence of the pure Harrison-Zel'dovich 
spectrum disappears. Notice that,  $\sigma^2$, and with it $\xi_c$ strongly depend 
on the smoothing scale. Therefore, although the Kaiser scheme leads to linear 
biasing of the non damped CDM power spectrum on small scales, once we remove 
the small scale power, the expected non-linear biasing on small scales re-appears.

If we choose a  very small damping scale, like e.g. 1pc or even 1kpc, the effect of smoothing
is not relevant in the observational window from $0.0003h{\rm Mpc}^{-1}<k<10h{\rm Mpc}^{-1}$.
As can be seen in Fig.~\ref{PSR}, damping is only relevant on scales $k\gsim R^{-1}$.
 
 It might be surprising at first to see in Fig.~\ref{PSR}  that the biased power spectra from 
larger damping scales are higher at given biasing parameter $\nu$. But this comes 
simply from the following fact: the larger the damping scale the 
smaller becomes the variance $\sigma$. For fixed $\nu$ the probability of  
$\de(\bx)/\si> \nu$ then increases 
and with it the correlation function $\xi_\nu$ and the power spectrum $P_\nu$. 
Alternatively, this can be understood by the increase of the normalized 
correlation function $\xi_c = \xi/\si^2$ on which $\xi_\nu$ depends 
monotonically. Remember however, as stressed above, this 
amplitude is not physical but simply a consequence of our normalization
in Eq.~(\ref{3}),  $\theta_\nu \le 1$.

In real observations, the {\em galaxy} correlation function, i.e. the biased  correlation function,
has to be smoothed over some 
scale $R$ e.g. with a Gaussian window function $W(r/R)$ so that
\be
P_\nu^{gal}=P_\nu(k,R)=|\widehat W(kR)|^2P_\nu(k).
\ee
In Fig.~\ref{Pnusmoothed} we show the corresponding results for 
the smoothed biased power spectrum. We see that the linear biasing on small scales disappears, but that the large-scale plateau remains because for scales larger than the smoothing radius the corresponding spectra are unaffected so that the non-linear biasing on large scales that we have found in the previous cases remain here as well.

\subsection{The baryon acoustic oscillations}

\begin{figure*}
\epsfig{width=17cm, file=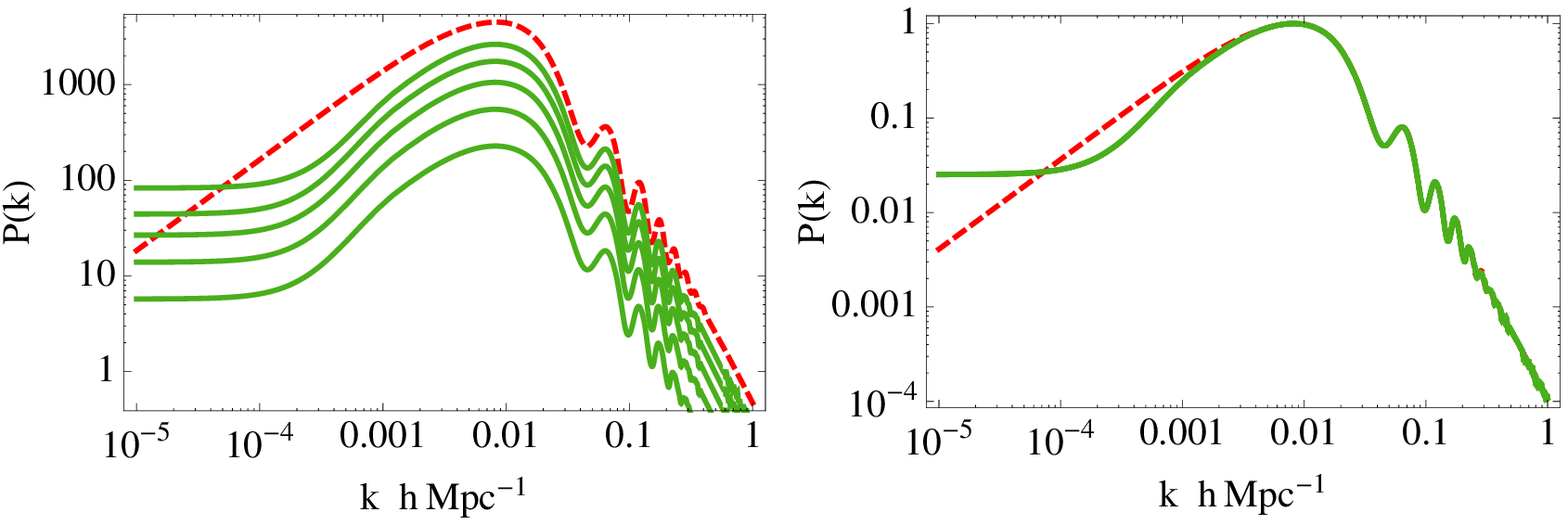}
\caption{Biasing in the presence of baryons. For this plot we choose 
$\Omega_bh^2=0.025$, $\Omega_Mh^2=0.05$ and, $h=0.5$. With this too large ratio
of $\Om_b/\Om_M$ the BAO's are well visible. In the left panel we show the 
underlying $\La$CDM power spectrum (red dashed line) that includes the effects 
of baryons and the corresponding biased power spectra (green solid lines) for 
$\nu=5$, $4$, $3$, $2$ and $1$ from top to bottom. In the right panel we have 
normalized the power spectra to their maximum value that clearly shows the 
two effects explained in the main text.}\label{PSBAO}
\vspace{0.3cm}
 \epsfig{width=17cm, file=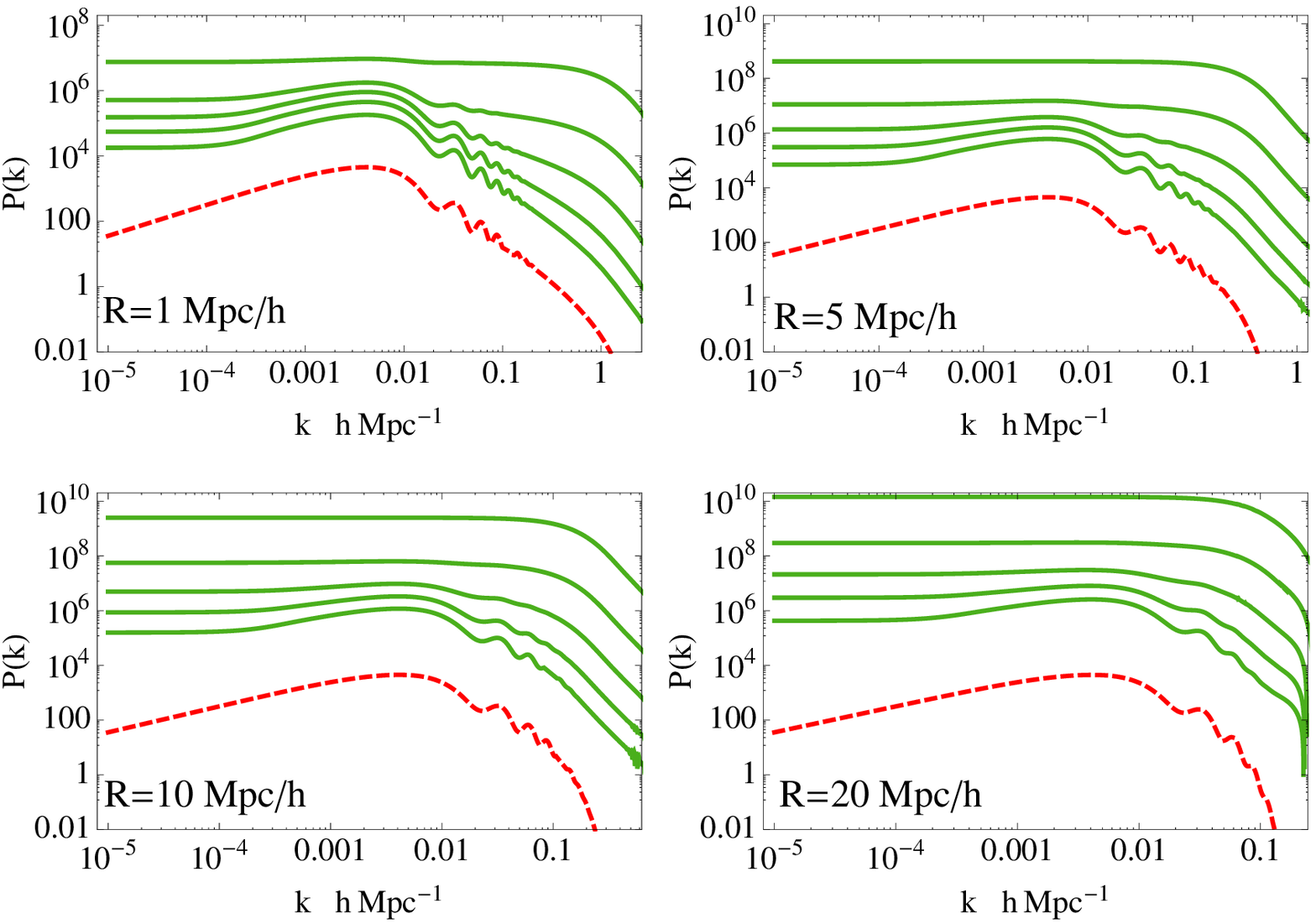}
\caption{The parameters are like for Fig.~\ref{PSBAO}. We show the underlying 
$\La$CDM power spectrum smoothed with a Gaussian window function (red dashed 
line) and the corresponding biased power spectra (green solid lines) for 
$\nu=5$, $4$, $3$, $2$ and $1$ from top to bottom for the different smoothing 
scales indicated in the panels.}\label{BAOsmooth}
\end{figure*}

\begin{figure*}
{ \epsfxsize=17cm \epsfbox{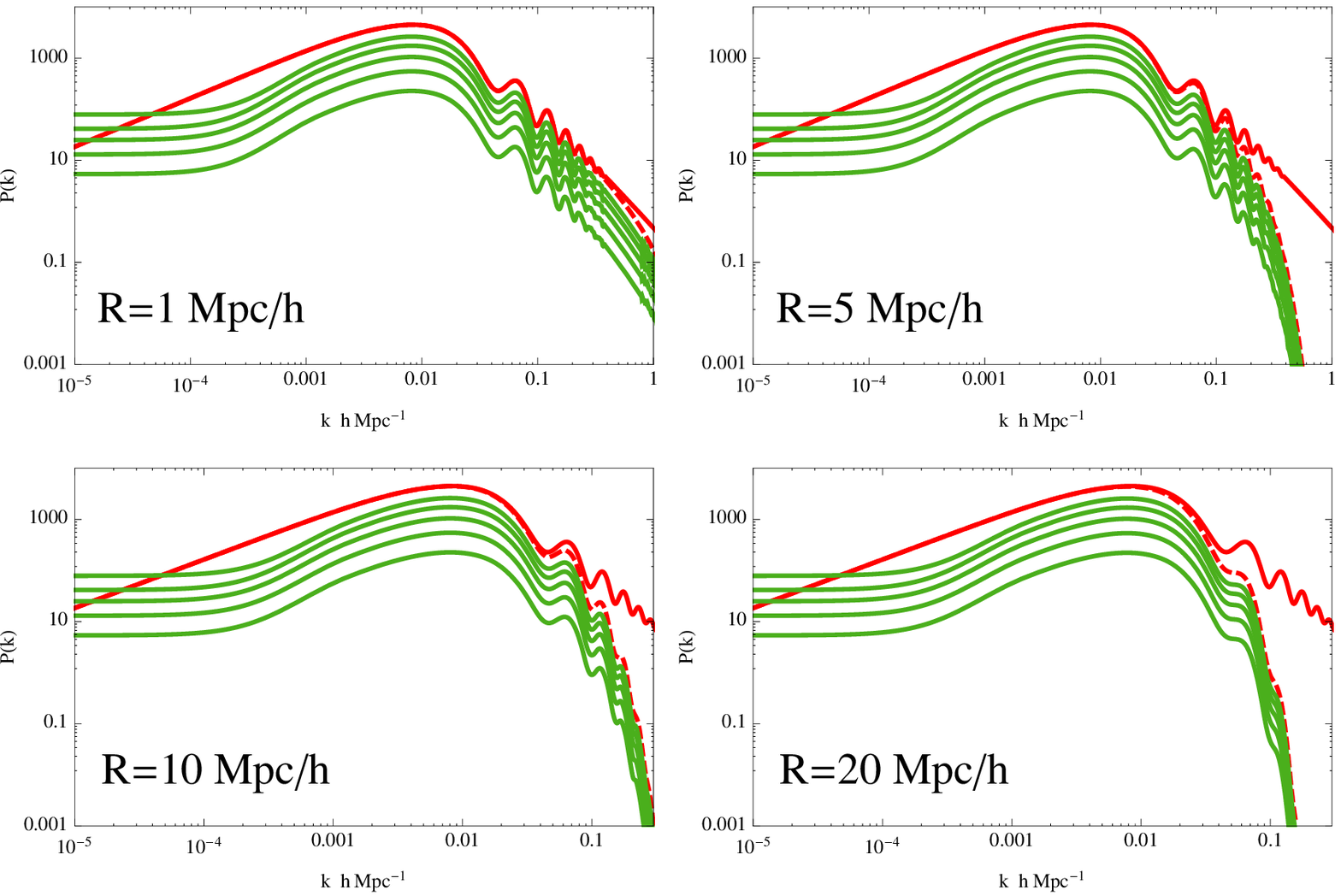}}
\caption{In this plot we show the biased power spectra after applying smoothing. The red line corresponds to the underlying power spectrum, the red-dashed line is the smoothed underlying power spectrum and the green lines show the biased power spectra after applying smoothing for $\nu=5,4,3,2,1$ from top to bottom.}
\label{PnuBAOsmooth}
\end{figure*}

We now take into account baryons and study the effects of the considered 
biasing scheme on the BAO's. For that, we shall  use the fitting formula given 
in~\cite{BAOPS} for the transfer function. Note that the effect of introducing 
baryons is not only the appearance of the BAO's, but also a suppression 
on small scales (Silk damping~\cite{mybook}). Also here, we use the linear transfer function.

In Fig. \ref{PSBAO} we show our results for this case.  In the plot 
we use a too small value of $\Om_M=2\Om_b=0.1$ and $h=0.5$ in order
to make the BAO's well visible. We have performed the calculation also for 
realistic values of the cosmological parameters and found the same results:
 the biased power spectrum flattens at the largest scales whereas the 
shape of the small scales of the spectrum, relevant for BAO's, remains 
unaffected and, 
therefore, the BAO's are not altered. Once again, this becomes more apparent 
when we normalize the spectra to their maximum value as shown in the right 
panel of Fig.~\ref{PSBAO}. As before, this result implies that the biased 
power spectra can be factorized as $P_\nu(k)=b_1(\nu)P_1(k)$ with $b_1(\nu)$ 
the function given in (\ref{b1}) and $P_1$ the universal shape of the 
biased power spectrum (which of course is different from the one 
without baryons). Hence Silk damping is not sufficient to alter the linear amplification
of the power spectrum on small scales. However, we do find a difference here with respect to the 
case without baryons: the power spectrum flattens at a larger scale  and 
 its decay towards the white noise plateau on large scales is even
steeper than the $P(k)\propto k^n$ behaviour of the underlying power spectrum. This 
difference can only be due to the different shape of the underlying power spectrum on 
small scales, the baryon acoustic oscillation and Silk damping. It is, however not clear 
whether this finding survives in a more realistic biasing scheme.

Also for this case, we have studied the effect of small scale damping of the underlying 
power spectrum with a Gaussian  window function. The corresponding results are 
shown in Fig.~\ref{BAOsmooth}. 
As before, the shape of the damped power spectrum is not conserved after 
biasing and the turnover tends to disappear for high values of the biasing 
parameter $\nu$ and for large smoothing scale. It is interesting to note that, 
unlike in the non-smoothed case, the BAO's are now distorted by biasing and they
are completely washed out for high values of $\nu$. We find interesting 
that smoothing over a scale of only $1h^{-1}$Mpc already has such a significant
effect on the power spectrum on large scales. As in the previous section without 
BAO's, we have linear biasing of the power 
spectrum on small scales which becomes non-linear once we introduce smoothing. 
Again, the reason for this is the high value of $\sigma^2$ of the non-smoothed 
power spectrum which is regularized by smoothing.

Finally, in order to compare to realistic observations of the galaxy distribution, we have 
considered the case of introducing smoothing on the biased power spectra. The 
corresponding results are shown in Fig. \ref{PnuBAOsmooth}. As in the previous 
section, the non-linear biasing on large scales remains unaltered because, as commented 
above, the scales larger than the smoothing scale are not affected. It is interesting to 
note however that  on scales comparable to the smoothing scale and smaller, the 
smoothed biased power spectra resemble the smoothed underlying power spectrum 
at small scales.

\section{Conclusions}\label{s:con}

In this work we have studied the effects of a simple biasing scheme on the 
matter power spectrum which has been considered before in Ref.~\cite{sylos}. 
However, we have used more realistic underlying power spectra, unlike in 
Ref.~\cite{sylos} where an unrealistic cutoff is introduced. We have studied both, the 
effect of such a cutoff and  the modification of the BAO's due to biasing.

First, we have used the approximate BBKS power spectrum, which neglects
baryons and the non-linear contribution, and we have found that the biased power spectrum is modified by two 
effects: a distortion at large scales where the power spectrum flattens and 
a vertical shift given by (\ref{b1}).  Contrary to the standard claim, in the 
power spectrum there is linear bias on small scales, large $k$,
$k_{\rm eq}< k \lsim 10^5$Mpc$^{-1}$ and a non-linear ($k$-dependent)
modification of the power spectrum on large scales. The linear biasing on 
small scales is due to the  UV divergence of the $n=1$ 
Harrison-Zel'dovich spectrum, that makes $\xi_c\ll 1$ already at $r=10^{-5}$Mpc
for the realistic value of $n=0.96$ and, therefore, the linear approximation is valid 
already above this scale. We have also studied the effect of biasing on a power 
spectrum  where small scale power  is damped
as e.g. in scenarios with warm dark matter. In that case, the biased power spectrum 
becomes distorted on all scales.  The UV-divergence is no longer present and 
the turnover on large scales tends to disappear for high values of the 
biasing parameter or large smoothing scale. Already for a smoothing scale of 
$1h^{-1}$Mpc, the turnover on large scales nearly disappears for $\nu>2$. 
If the damping scale is very small, $R\lsim 1$kpc, smoothing has no effect 
on the power spectrum on the scales considered here.

In real observations, there is of course also smoothing of the galaxy power spectrum 
due to finite resolution. We have introduced this smoothing in the biased power 
spectrum and we have seen that the linear amplification on small scales also 
disappears and the power spectra flatten on large scales. However, in this case 
the turnover can be visible.

Finally, we have included baryons and studied the possible effects on the 
BAO's. We have found the same features as 
in the case without baryons:  the biased power spectrum 
flattens at large scales, and it is linearly re-scaled by the constant factor 
$b_1(\nu)$ at small scales. There are no effects on the BAO's. 
They are not smeared out by Kaiser biasing.  
However,  damping of the underlying power spectrum distorts the biased 
power spectrum at all scales. For a damping scale of 
$R\simeq 1$Mpc/h and a bias parameter $\nu\lsim 2$, the BAO's remain intact.
However, they completely disappear for high values of the biasing parameter, 
$\nu> 3$ and large damping scale, $R>5$Mpc/h. Again, we have also introduced smoothing 
of the biased power spectra in order to study the effects in real observations. We have seen that BAO's are not affected if the smoothing scale is small enough.
However, we cannot definitively decide
on the modification of the BAO's by biasing. The result depends on the 
details of the underlying biasing scheme and is not robust.  

Our findings show that, for the biasing scheme considered in this work, the 
$P\propto k^n$ slope is in principle observable for nearly a decade before 
the 'white noise' contribution sets in. When we introduce small scale damping of 
the underlying power spectrum
the situation becomes worse and the slope can become unobservable, even the 
turn over may disappear. Finally, realistic observations require  smoothing of the 
biased power spectra so we have also studied this and we have seen that the 
turnover is not affected by smoothing on small scales.  This result is less pessimistic 
than in Ref.~\cite{sylos}, but it has to be checked with a more realistic 
biasing scheme that will be considered in a future project.

{\bf Acknowledgments:} 
We thank Vincent Desjacques, Martin Kunz  and Francesco Sylos Labini for useful 
discussions and comments.
JBJ acknowledges support from MICINN (Spain) project numbers
FIS 2008-01323 and FPA 2008-00592, MEC grant BES-2006-12059 
and MICINN Consolider-Ingenio MULTIDARK CSD2009-00064 and wishes to thank  
the Theoretical Physics Department of the Geneva University for hospitality. 
This work is supported by the Swiss National Science Foundation. 

\end{document}